\documentclass[aps,pra,showpacs,superscriptaddress,twocolumn,longbibliography]{revtex4-1}
\usepackage{amsfonts}
\usepackage{subfigure}
\usepackage{amsmath}
\usepackage{txfonts}
\usepackage{amssymb}
\usepackage{amsbsy}
\usepackage{epsfig}
\usepackage{graphicx}
\usepackage{epstopdf}

\begin{document}

\title{The topological counterparts of non-Hermitian SSH models}
\date{\today }
\email{csliu@ysu.edu.cn}
\author{Y. Z. Han, J. S. Liu, C. S. Liu}

\begin{abstract}
The breakdown of the conventional bulk-boundary correspondence due to
non-Hermitian skin effect leads to the non-Bloch bulk-boundary
correspondence in the generalized Brillouin zone. Inspired by the case of
the equivalence between the non-reciprocal hopping and imaginary gauge
field, we propose a method to construct the topological equivalent models of
the non-Hermitian dimerized lattices with the similarity transformations.
The idea of the constructions is from that the imaginary magnetic flux
vanishes under the open boundary condition and the period boundary spectra
can be well approximated by open boundary spectra. As an illustration, we
apply this approach to several representative non-Hermitian SSH models,
efficiently obtaining topological invariants in analytic form defined in the
conventional Bloch bands. The method gives an alternative way to study the
topological properties of non-Hermitian system.
\end{abstract}

\maketitle

\affiliation{Hebei Key Laboratory of Microstructural Material Physics,
School of Science, Yanshan University, Qinhuangdao, 066004, China}

\section{Introduction}

The nonconservative phenomena existing widely in natural and artificial
systems, e.g., open system coupled to energy or particle sources, or driven
by external fields can be modeled by the effective non-Hermitian
Hamiltonians \cite{Rotter_2009, PhysRevLett.85.2478, Bender_2007,
Choi2010Quasieigenstate, Moiseyev, PhysRevA.85.032111, Lu2014Topological,
Rotter_2015}. The non-Hermitian systems have exhibited rich exotic
characteristics without Hermitian counterparts \cite{PhysRevB.56.8651,
PhysRevLett.77.570, PhysRevB.84.205128, PhysRevA.89.062102, YUCE20151213,
PhysRevLett.116.133903, PhysRevLett.118.040401, PhysRevLett.120.146402,
PhysRevB.97.121401, Xiong_2018, PhysRevLett.121.026808, PhysRevX.8.031079,
PhysRevB.99.125103, PhysRevX.9.041015, PhysRevLett.123.066405}.
%The more types of symmetries lead to topological classification beyond the standard ten classes.
The prominent character is the breakdown of the usual bulk-boundary
correspondence in the nonreciprocal lattices due to the non-Hermitian skin
effect \cite{Xiong_2018, PhysRevB.97.121401, PhysRevLett.121.026808,
PhysRevLett.121.086803, PhysRevLett.121.136802, PhysRevLett.123.066404,
PhysRevLett.123.170401, PhysRevB.99.201103, PhysRevLett.123.246801,
PhysRevLett.125.126402, PhysRevB.99.081103, PhysRevB.101.020201,
PhysRevB.102.085151}. The non-Hermitian skin effect induces all the bulk
states are localized at the boundaries of the system and are
indistinguishable from the topological edge states. The novel topological
invariants are needed to character topological properties. The pioneering
research is the proposal of the generalized Brillouin zone (GBZ) which
recovers the correspondence between the winding number based on complex
energy with periodic boundaries and the existence of zero modes with open
boundaries \cite{PhysRevLett.121.086803}. %Up to now, most studies of
%the non-Hermitian phenomena are general to solve the non-Hermitian
%Hamiltonian directly.
The calculation of GBZ becomes an important issue and has drawn extensive
attentions recently \cite{PhysRevLett.125.126402, PhysRevLett.125.186802,
PhysRevLett.125.226402}.

It is well known that there is a Hermitian counterpart for a non-Hermitian
Hamiltonian within the symmetry-unbroken region, in which the two
Hamiltonians have an identical fully real spectrum \cite%
{PhysRevLett.80.5243, doi:10.1063/1.1418246, PhysRevB.97.115436}. This
allows us to find the nontrivial states of the non-Hermitian Hamiltonian
with the open boundary condition from its Hermitian counterpart. Here, a
natural question that arises in this topics is whether there is a partner of
non-Hermitian system with non-Hermitian skin effect that share the same
topological phase diagrams. As such, the topological invariant of the
original model can be obtained from its partner, which can be calculated in
an easier way. In particular, the topological invariant of their partner may
have the analytical form which is important to understand the whole
properties of non-Hermitian system.

In general, the non-Hermitian skin effect comes from the non-reciprocal
hopping of the lattice. This effect can be realized experimentally in atomic
systems by laser assisted spin-selective dissipations \cite{Lapp_2019,
Nat_commun_855_1}. As the case studied in Ref. [%
\onlinecite{PhysRevLett.116.133903}] and [\onlinecite{PhysRevB.97.121401}],
the model with balanced hopping and only gains and losses is mathematically
equivalent to the non-Hermitian Su-Schrieffer-Heeger (SSH) model with the
imbalance hopping when Pauli matrix $\sigma_z$ is replaced by $\sigma_y$
although the original model can not be interpreted by SSH model. The recent
study provides the conditions under which on-site dissipations can induce
non-Hermitian skin modes \cite{PhysRevLett.125.186802}.

The asymmetric couplings are equivalent to an imaginary gauge field applying
to the lattice \cite{PhysRevLett.77.570, PhysRevB.99.081103}.
%As we know, the magnetic filed $\textbf{B}$ enters the Schr\"{o}rding equation through the vector potential $\textbf{A}$.
%The periodic boundary spectra under a uniform and perpendicular image magnetic field can be obtained from that of without %magnetic field case by replacing the wave vector $\textbf{A}\rightarrow\textbf{k}+i\textbf{A}$.
Under the periodical boundary condition, the imaginary gauge field enclosed
in an area induces a nonzero imaginary magnetic flux, which is non-Hermitian
Aharonov-Bohm effect as the Hermitian case \cite{PhysRevB.99.081103}. As
pioneered by Yang and Lee in 1952, quantum phase transition can be driven by
a complex external parameter \cite{PhysRev.87.404, PhysRev.87.410}. The
nonzero imaginary magnetic flux breaks the conventional bulk-boundary
correspondence and leads to a topological phase transition. Under the open
boundary condition (OBC) however, the imaginary gauge field is not enclosed
in an area and the imaginary magnetic flux vanishes \cite{PhysRevB.99.081103}%
. Although the OBC breaks the translational symmetry, according to the
bulk-boundary correspondence in the long chain case, the boundary scattering
can still be regarded as a perturbation whether in Hermit case or not due to
the introduction of the GBZ. This gives a strong hint that the Bloch
Hamiltonian can be well approximated by the bulk Hamiltonian under the OBC.
Following this route, non-Hermitian Bulk-boundary correspondence of the
non-Hermitian SSH model without the $t_3$ was recovered after getting rid of
the effective imaginary gauge \cite{PhysRevB.99.081103}. A key step is
transforming the non-Hermitian terms to the Bloch phase factor with a
similarity transformation. This step can also be realized under the OBC
through another similarity transformation \cite{PhysRevLett.121.086803}.
After the similarity transformation, the non-Hermitian SSH model is
transformed to its topological equivalent model which leads to understand
the nontrivial topological phase easily.

A natural issue is whether the method can be extended to construct the
partner of the general non-Hermitian model with the asymmetric coupling
terms. Motivated by the above considerations, we develop this method to
construct the partners of several non-Hermitian SSH models. As is shown in the
following studies, the central tenet of the constructing is building the
relationship between non-Hermitian skin effect and imaginary equivalent gauge field.
The topological invariants based on band-theory are effective to predict the
topological nontrivial states. The remainder of this paper is organized as
follows. In Sec. \ref{The non-Hermitian SSH models}, we present the
Hamiltonians of non-Hermitian SSH model and its various equivalent models.
We show how the non-Hermitian SSH model is transformed to its equivalent
models by the similarity transformations and how the non-Hermitian skin effect is eliminated due to the offset of the imaginary gauge field.
In subsection \ref{The partner
model}, the effectively gauge field is find to construct a partner model of
the non-Hermitian SSH model by a similarity transformation. Due to the
topologically equivalence of the two models, we study the topological phase
transitions with the partner models in subsection \ref{The topological
invariant}. Inspirited by the consistence of the QPTs with the numerical
method, we further apply the method to
study the non-Hermitian SSH model with spin-orbit coupling in Sec.
\ref{The non-Hermitian SSH model with spin-orbit coupling} and
the topological defect states of the non-Hermitian
SSH model in Sec. \ref{The topological defect on non-Hermitian SSH model}.
Finally, a summary and discussion are given in Sec. \ref{Summary}.

%\section{Models and Methods}

%\label{Models and Methods}

\section{The non-Hermitian SSH models}

\label{The non-Hermitian SSH models}

The non-Hermitian SSH model is described Bloch Hamiltonian $%
H_{k}=\psi_{k}^{\dag }h_{k}\psi _{k}$ with $h_{k}=\left( \vec{d}-i\vec{\Gamma%
}\right)\cdot \vec{\sigma}$, where $\vec{\sigma}=\left( \sigma
_{x},\sigma_{y},\sigma _{z}\right)$ is the Pauli matrix for spin-1/2 and $%
d_{x}=t_{1}+(t_{2}+t_{3})\cos k$, $d_{y}=(t_{2}-t_{3})\sin k$ \cite%
{PhysRevLett.42.1698, PhysRevB.102.161101}. The Numb wavefunction $\psi
_{k}^{\dag }=(a_{k}^{\dag },b_{k}^{\dag })$. When Pauli matrix $\sigma_y$ is
replaced by $\sigma_z$ and setting $t_1 = M+4B$, $t_2+t_3=2B$ and $%
t_2-t_3=2A $, the generalized SSH model $h_{k}$ can be mapped to 1D
Creutz-type model \cite{PhysRevLett.83.2636} which is the dimensional
reduced BHZ model \cite{Bernevig_2006, Guo:2016aa}.
When Pauli matrix $\sigma_x$ is
replaced by $\sigma_z$ and replacing $t_1 = -\mu$, $t_2+t_3=-2t$ and $%
t_2-t_3=-2\Delta$, the generalized SSH model $h_{k}$ can be mapped to 1D
Kitaev model model \cite{Kitaev_2001, Guo:2016aa}.
The non-Hermiticity
comes from the term $\vec{\Gamma}$. When $\vec{\Gamma}=\left(0,0,\gamma/2%
\right)$, the non- Hermiticity doesn't change the topological phase
transition and the existence of the edge state due to the
pseudo-anti-Hermiticity protection \cite{PhysRevA.98.052116,
PhysRevLett.121.026808, PhysRevB.100.045141}. If the imaginary part $\vec{%
\Gamma}=\left(\gamma/2,0,0\right)$, the inversion symmetry of $h_k$ ensures
non-existence of the non-Hermitian skin effect and the validity of
conventional bulk-boundary correspondence. The nontrivial topology is
established from the linking geometry of the vector fields \cite%
{PhysRevB.102.161101}. We study the case of $\vec{\Gamma}=\left(0,\gamma/2,0%
\right)$. The non-Hermitian SSH model and its equivalent two-leg ladder
model are pictorially shown in Fig. \ref{The non-Hermitian SSH model} (a)
and (b). The Bloch Hamiltonian reads
\begin{equation}
h_{k}=d_{x}\sigma _{x}+(d_{y}+i\frac{\gamma }{2})\sigma _{y}
\label{The non-Hermitian SSH model}
\end{equation}%
where $\sigma _{x,y}$ are the Pauli matrices The asymmetric intracell
coupling amplitude $t_{1}\pm \gamma /2$ can be realized in open classical
and quantum systems with gain and loss \cite{Lapp_2019, Nat_commun_855_1,
2020NatPh..16..761X, 2020NatPh..16..747H}. The model has a chiral symmetry $%
\sigma _{z}^{-1}h_{k}\sigma _{z}=-h_{k}$, which ensures that the eigenvalues
appear in $(E,-E)$ pairs.

\begin{figure}[tbp]
\begin{center}
\includegraphics[width=8cm]{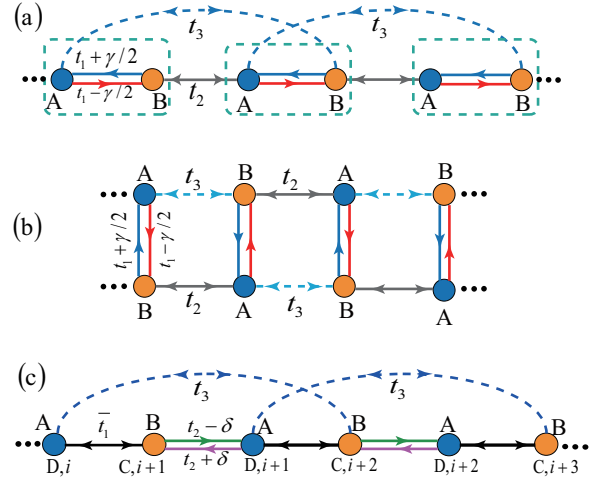}
\end{center}
\caption{(a) Non-Hermitian SSH model and (b) its equivalent two-leg ladder
model. (c) Schematic of the mapping between the non-Hermitian SSH model in
Eq. (\protect\ref{The non-Hermitian SSH model}) and the extended
non-Hermitian SSH model in Eq. (\protect\ref{The extended non-Hermitian SSH
model}).}
\label{Non_Hermit_SSH}
\end{figure}

There are various equivalent models for this model. For example, a
mathematically equivalent model is studied in Ref. [%
\onlinecite{PhysRevB.84.205128}]. This model can be obtained by a similarity
transformation $\bar{H}=S_{1}^{-1}HS_{1}$ with a diagonal matrix $S_{1}$
whose diagonal elements are judiciously chosen as \cite%
{PhysRevLett.121.086803}
\begin{equation}
S_{1}=\{1,r_{1},r_{1},r_{1}^{2},r_{1}^{2},\cdots
,r_{1}^{L/2-1},r_{1}^{L/2-1},r_{1}^{L}\}
\label{the transformation matrix S1}
\end{equation}
here $r_{1}=\sqrt{\left\vert \left( t_{1}-\gamma /2\right) /\left(
t_{1}+\gamma /2\right) \right\vert }$. Under the similarity transformation,
the non-Hermiticity in $t_{1}$ term is transformed to $t_{3}$ term where $%
t_{3}$ term becomes $t_{3}r_{1}^{2}$ and $t_{3}/r_{1}^{2}$. The wavefunction
$|\psi \rangle =(a_{1},b_{1},a_{2},b_{2},\cdots ,a_{L},b_{L})^{T}\ $\
becomes $|\bar{\psi}\rangle =S_{1}^{-1}|\psi \rangle $. The intracell
hopping $\bar{t}_{1}=\sqrt{|t_{1}^{2}-\gamma ^{2}/4|}$ and the $t_{2}$ term
remains unchanged.

The extended non-Hermitian SSH discussed in Ref. [%
\onlinecite{PhysRevA.97.052115}] can be obtained by a similarity
transformation $\tilde{H}=S_{2}^{-1}\bar{H}S_{2}$ with a diagonal matrix
\begin{equation}
S_{2}=\text{diag}\{r_{2},r_{2},r_{2}^{2},r_{2}^{2},\cdots
,r_{2}^{L/2},r_{2}^{L/2}\}  \label{the transformation matrix S2}
\end{equation}%
here $r_{2}=r_{1}^{-2}$. After this transformation shown in Fig. \ref{The
non-Hermitian SSH model} (c), the non-Hermiticity is transformed from $t_{3}$
term to $t_{2}$ term where $t_{2}$ term becomes $t_{2}/r_{2}$ and $%
t_{2}r_{2} $. The wavefunction $|\tilde{\psi}\rangle =S_{2}^{-1}|\bar{\psi}%
\rangle $ and the intracell hopping $\bar{t}_{1}$ remains unchanged. By
relabeling the sites A$\rightarrow $D and B$\rightarrow $C, the model can be
mapped to an extended non-Hermitian SSH model. The Hamiltonian in momentum
space takes the form
\begin{equation*}
\tilde{H}_{k}=\sum_{k}\tilde{\psi}_{k}^{\dagger }\tilde{h}_{k}\tilde{\psi}%
_{k},
\end{equation*}%
where $\tilde{\psi}_{k}=(c_{k},d_{k})^{T}$ and
\begin{equation}
\tilde{h}_{k}=\left(
\begin{array}{cc}
0 & t-\delta +t^{\prime }e^{ik}+\Delta e^{-2ik} \\
t+\delta +t^{\prime }e^{ik}+\Delta e^{2ik} & 0%
\end{array}%
\right) .  \label{The extended non-Hermitian SSH model}
\end{equation}%
In the mapping, the we have replaced $\Delta =t_{3}$, $t^{\prime }=\bar{t}%
_{1}$, $t-\delta =t_{2}/r_{2}$ and $t+\delta =t_{2}r_{2}$.

The non-Hermitian SSH model in Eq. (\ref{The non-Hermitian SSH model}) is
also topological equivalence to the model discussed in Ref. [%
\onlinecite{PhysRevLett.123.066404}] where the non-Hermiticity occurs in $%
t_1 $, $t_2$ and $t_3$ terms. One can first do the inverse similarity
transformation in Eq. (\ref{the transformation matrix S2}). The
non-Hermiticity in $t_2$ term is transferred to $t_3$ term. Then further
doing the inverse similarity transformation in Eq. (\ref{the transformation
matrix S1}), the non-Hermiticity in $t_3$ term is transferred to $t_1$ term.
After doing the two similarity transformations, the non-Hermiticity in $t_2$
and $t_3$ terms disappear and the asymmetric intercell coupling is
transformed to symmetrical. The intercell coupling relating to $t_1$ term
remains asymmetrical.

\section{The partner model and topological invariant}

\label{The partner model and topological invariant}

\subsection{The partner model}

\label{The partner model}

\begin{figure}[tbp]
\begin{center}
\includegraphics[width=8cm]{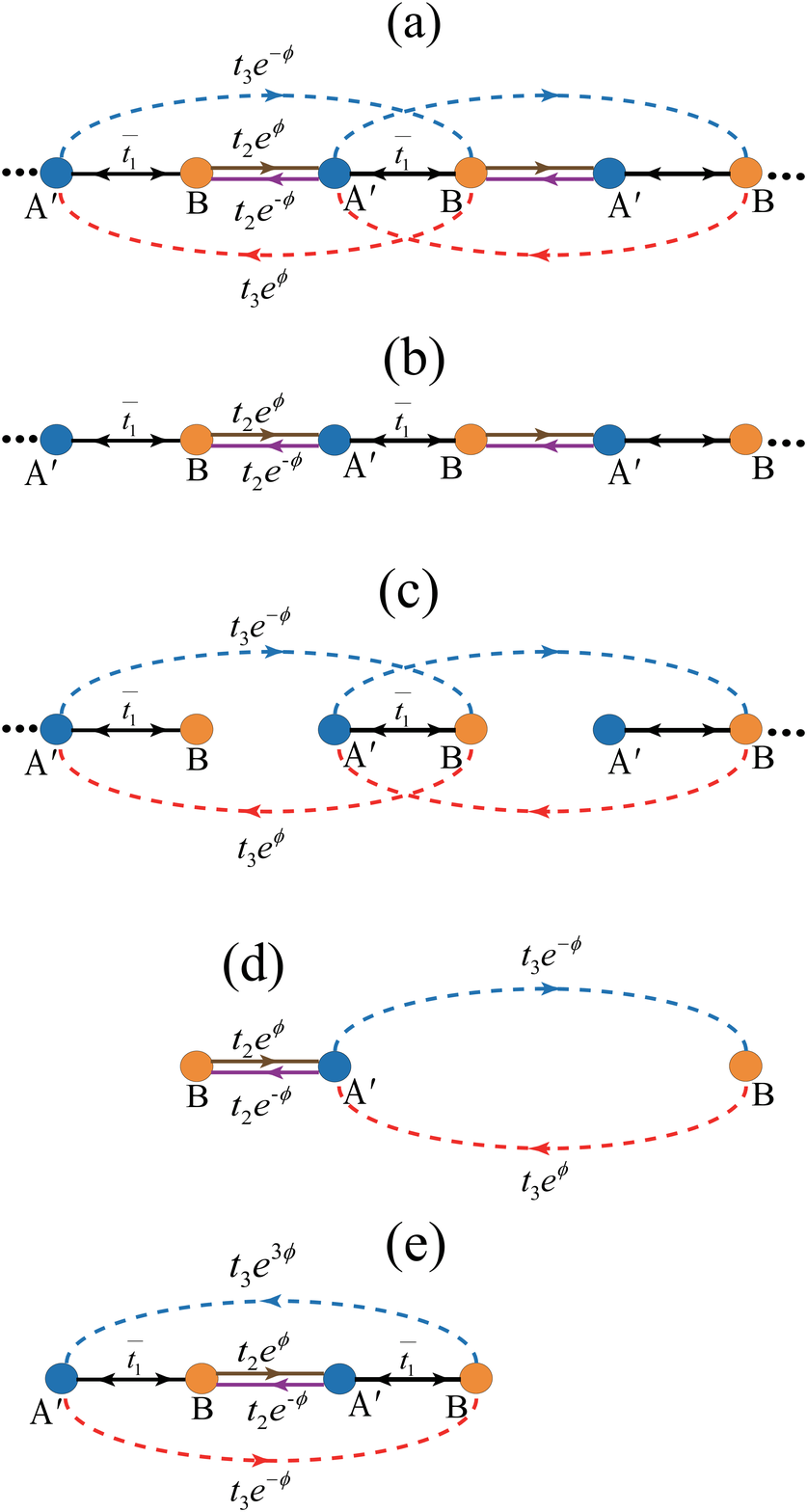}
\end{center}
\caption{(a) Schematic of the imaginary gauge field in Eq. (\protect\ref{The
non-Hermitian SSH model I}) when $t_2 \neq 0$ and $t_3 \neq 0$. Illustration
of the non-Hermitian skin effect when (b) $t_2 \neq 0$, $t_3 0 0$ and (c) $%
t_2 = 0$, $t_3 \neq 0$. (d) Illustration of the effective imaginary gauge
field of the non-Hermitian SSH model in Eq. (\protect\ref{The non-Hermitian
SSH model I}). (e) Illustration of the non-Hermitian skin effective of the
partner model in Eq. (\protect\ref{the partner Hamiltonian}).}
\label{imaginary gauge_field}
\end{figure}

The asymmetric couplings in $t_{1}$ term of Eq. (\ref{The non-Hermitian SSH
model}) can be expressed as a symmetric coupling $\bar{t}_{1}$ with phase
factor of amplification/attenuation $e^{\pm \phi }$, i.e. $t_{1}\pm \gamma
/2=\bar{t}_{1}e^{\pm \phi }$. Under the basis $\psi _{k}^{\prime \dag
}=\{e^{-\phi }a_{k}^{\dagger },b_{k}^{\dagger }\}=\{a_{k}^{\prime
}{}^{\dagger },b_{k}^{\dagger }\}$, the Hamiltonian $H_{k}$ can be rewritten
in the form of
\begin{eqnarray}
H_{k} &=&\psi _{k}^{\prime \dag }\hat{h}_{k}\psi _{k}^{\prime }=\left[ \bar{t%
}_{1}+t_{2}e^{\phi }e^{-ik}+t_{3}e^{\phi }e^{ik}\right] a_{k}^{\prime
}{}^{\dagger }b_{k}  \notag \\
&&+\left[ \bar{t}_{1}+t_{2}e^{-\phi }e^{ik}+t_{3}e^{-\phi }e^{-ik}\right]
b_{k}^{\dagger }a_{k}^{\prime }{}.  \label{The non-Hermitian SSH model I}
\end{eqnarray}%
With the similarity transformation, the AB sublattice becomes A$^{\prime}$B
sublattice and the asymmetric coupling is transformed from $t_{1}$ term to $%
t_{2}$ and $t_{3}$ terms shown in Fig. \ref{imaginary gauge_field} (a).
Here, the site label A is replaced by A$^{\prime}$ which describes the
annihilation operator $a$ of A site is transformed to $a^{\prime}$.

Assuming $t_{1}$, $t_{2}$, $t_{3}$ and $\phi $ to be greater than zero, the
non-Hermitian skin effect in model (\ref{The non-Hermitian SSH model I}) can
been analysed in real space as follows. When $t_{2}\neq 0$ and $t_{3}=0$
shown in \ref{imaginary gauge_field} (b), the model (\ref{The non-Hermitian
SSH model I}) is reduced to the general non-Hermitian SSH model. The hopping
$t_{2}e^{\phi }$ from B sites to A$^{\prime }$ sites is larger than that $%
t_{2}e^{-\phi }$ from A$^{\prime }$ to B sites. The asymmetry hopping leads
to the particles tend to right side. However, in the case of $t_{3}\neq 0$
and $t_{2}=0$ shown in \ref{imaginary gauge_field} (c), the model (\ref{The
non-Hermitian SSH model I}) is also reduced to the general non-Hermitian SSH
model where the hopping $t_{3}e^{\phi }$ from A$^{\prime }$ sites to B sites
is larger than that $t_{3}e^{-\phi }$ from B to A$^{\prime }$ sites which
induces the particles tend to left side. For the case $t_{2}=t_{3}\neq 0$,
the Hamiltonian in Eq. (\ref{The non-Hermitian SSH model I}) has the $%
\mathcal{PT}$ symmetry and the two non-Hermitian skin effects cancel.
Therefore, we conclude that there exist a non-Hermitian skin effect when $%
t_{2}\neq t_{3}$. When $t_{2}>t_{3}$, the non-Hermitian skin effect is
governed by the $t_{2}$ term and all the states are localized in the right
edge. When $t_{2}<t_{3}$, $t_{3}$ term dominates the non-Hermitian skin
effect which leads the all the states localized in the right edge.

The amplification and attenuation factors $e^{\pm \phi }$ in Eq. (\ref{The
non-Hermitian SSH model I}) indicate an imaginary gauge field $\phi$
applying to the lattice. The effective imaginary gauge field can be analysed
in moment space as follows. In the unit cell shown in Fig. \ref{imaginary
gauge_field} (d), two channels are provided for a particle tunneling from A$%
^{\prime }$ to B with different hopping amplitude. It indicates that a
particle tunneling from A$^{\prime }$ site to B site will obtain a $e^{\phi}$
phase for the $t_2 $ and $t_3$ channels. The product of the above two terms
contribute the overall accumulated phase factor $e^{2\phi }$ which suggests
the enclosed imaginary gauge field be $2\phi $.

We therefore rewrite $\hat{h}_{k}$ in Eq. (\ref{The non-Hermitian SSH model
I}) as
\begin{eqnarray}
\check{h}_{k} &=&\left[ \bar{t}_{1}+t_{2}e^{-i\left( k+2i\phi \right)
}e^{-\phi }+t_{3}e^{i\left( k+2i\phi \right) }e^{3\phi }\right] \sigma _{+}
\label{The non-Hermitian SSH model II} \\
&&+\left[ \bar{t}_{1}+t_{2}e^{i\left( k+2i\phi \right) }e^{\phi
}+t_{3}e^{-i\left( k+2i\phi \right) }e^{-3\phi }\right] \sigma _{-},  \notag
\end{eqnarray}%
where $\sigma _{\pm }=\sigma _{x}\pm i\sigma _{y}$. Considering the terms $%
t_{2}e^{-i\left( k+2i\phi \right) }$ and $t_{3}e^{i\left( k+2i\phi \right) }$
in Eq. (\ref{The non-Hermitian SSH model II}), they are equivalent to an
imaginary magnetic field applying to the A$^{\prime }$B lattice. We take a
complex-valued wave vector $\tilde{k}\rightarrow k+2i\phi $ to describe
open-boundary eigenstates. In this replacement, the Hamiltonian (\ref{The
non-Hermitian SSH model II}) can be treated as parameter of $\tilde{k}$
which a deformation of the standard Brillouin zone. Accordingly, we define a
partner Hamiltonian as follows:
\begin{equation}
\check{h}_{\tilde{k}}=h_{+}\left( \tilde{k}\right) e^{i\theta _{+}\left(
\tilde{k}\right) }\sigma _{+}+h_{-}\left( \tilde{k}\right) e^{i\theta
_{-}\left( \tilde{k}\right) }\sigma _{-},  \label{the partner Hamiltonian}
\end{equation}%
which is also non-Hermitian Hamiltonian. $h_{+}( \tilde{k})$ and $\theta
_{+}( \tilde{k}) $ are the module and angle of complex function $\bar{t}%
_{1}+t_{2}e^{-i\tilde{k}}e^{-\phi }+t_{3}e^{i\tilde{k}}e^{3\phi }$. $h_{-}(
\tilde{k})$ and $\theta _{-}( \tilde{k}) $ are the module and angle of
complex function $\bar{t}_{1}+t_{2}e^{i\tilde{k}}e^{\phi }+t_{3}e^{-i\tilde{k%
}}e^{-3\phi }$.

The partner model is still a non-Hermitian model. The non-Hermitian $t_2$
and $t_3$ terms lead to the accumulation of real phase factor which suggests
the enclosed imaginary gauge field be $2\phi$. The $2\phi$ imaginary gauge
field suggests that the wave vector should be $\tilde{\tilde{k}}\rightarrow
\tilde{k}+2i\phi $. The model in Eq. (\ref{the partner Hamiltonian}) is
transformed back to the model in Eq. (\ref{The non-Hermitian SSH model I}).

The non-Hermitian skin effect in model (\ref{the partner Hamiltonian}) can
also been analysed in real space. In the unit cell shown in Fig. \ref%
{imaginary gauge_field}(e), $\pm\phi$ in the asymmetry hopping $t_2$ term is
equivalent to an imaginary magnetic field applied to the lattice with an
imaginary magnetic vector potential $i\phi$ alone $x$ direction. For the
asymmetry hopping $t_3$ term, $\pm 3\phi$ can be written as $\pm \phi (3a)$
here the lattice constant $a=1$. $3a$ indicates three sites have been
crossed when a particle hopping from B site to A$^\prime$ through the $t_3$
channel. It is also equivalent to an imaginary magnetic field applied to the
lattice with an imaginary magnetic vector potential $i\phi$. However, the
imaginary magnetic vector potential is alone the $-x$ direction. The two
imaginary magnetic vector potentials are cancelled and the imaginary
magnetic flux does not exist under the periodical boundary condition. No
non-Hermitian skin effect occurs in the model [Eq. (\ref{The non-Hermitian
SSH model II})] and wave vector is still a good quantum number.

The partner model in Eq. (\ref{the partner Hamiltonian}) is also effective
to the case $t_2 \neq 0$ and $t_3=0$. Replacing $\tilde{\tilde{k}}%
\rightarrow \tilde{k}-i\phi $, the partner model in Eq. (\ref{the partner
Hamiltonian}) is transformed the standard SSH model with the phase
transition point $\bar{t}_1 = t_2$. When $t_3 \neq 0$ and $t_2=0$, we can
replace $\tilde{\tilde{k}}\rightarrow \tilde{k}+ 3i\phi $. The partner model
in Eq. (\ref{the partner Hamiltonian}) is also transformed the standard SSH
model with the phase transition point $\tilde{t}_1 = t_3$. The above cases
have been studied in Ref. [\onlinecite{PhysRevLett.121.086803}] and [%
\onlinecite{PhysRevB.99.081103}].

Under the OBC, the partner Hamilton in Eq. (\ref{the partner Hamiltonian})
can also be obtained from the model in Eq. (\ref{The non-Hermitian SSH model}%
) by a similarity transformation $\bar{H}=S_{3}^{-1}HS_{3}$ directly. $%
S_{3}=S_{1}S_{2}$ is also a diagonal matrix where $S_{1}$ and $S_{2}$ are
given in Eq. (\ref{the transformation matrix S1}) and (\ref{the
transformation matrix S2}) with $r_{1}=r_{2}=e^{\phi }$. The similarity
transformation changes the eigen-states and don't change the its
eigen-energy. So the models in Eq. (\ref{The non-Hermitian SSH model}) Eq. (%
\ref{The non-Hermitian SSH model I}) and Eq. (\ref{the partner Hamiltonian})
are topological equvalent. As shown in Subsec. \ref{The topological
invariant}, the topological nontrivial phases can be charactered by the
winding number based on the Eq. (\ref{the partner Hamiltonian}).

\subsection{The topological invariant}

\label{The topological invariant}

According to the usual bulk-boundary-correspondence scenario, the chiral
edge states of an 1D open-boundary system should be determined by the
winding numbers which are closely related to the Zak phase across the
Brillouin zone \cite{PhysRevLett.62.2747}. For the non-Hermitian system with
chiral symmetry and the complex eigenvalues, the winding number of energy is
defined as a topological invariant \cite{PhysRevLett.118.040401,
PhysRevLett.120.146402, PhysRevA.98.052116}. As $\tilde{k}$ goes across the
generalized Brillouin zone, the winding number of energy $\nu _{E}$ is
defined as
\begin{equation}
\nu _{E}=\frac{1}{2\pi }\oint d\tilde{k}[\partial _{\tilde{k}}\arg \left(
E_{2}-E_{1}\right) ]  \label{the winding number of energy}
\end{equation}%
where the integral is also taken along a loop with $\tilde{k}$ from $0$ to $%
2\pi $. The eigenvalues of Hamiltonian (\ref{the partner Hamiltonian}) are
\begin{equation*}
E_{1,2}=\pm \sqrt{h_{+}\left( \tilde{k}\right) h_{-}\left( \tilde{k}\right) }%
\exp \left\{ i\left[ \theta _{+}\left( \tilde{k}\right) +\theta _{-}\left(
\tilde{k}\right) \right] /2\right\}
\end{equation*}%
which are smoothly continuous with $\tilde{k}$. $\nu _{E}$ is summation of
winding numbers of two winding vectors $h_{+}\left( \tilde{k}\right)
e^{i\theta _{+}\left( \tilde{k}\right) }$ and $h_{-}\left( \tilde{k}\right)
e^{i\theta _{-}\left( \tilde{k}\right) }$. In Hermitian systems, $\nu _{E}$
is always zero due to the real energy $E_{1,2}$.

The non-Bloch winding number can also be introduced with the
\textquotedblleft $Q$ matrix\textquotedblright\ \cite{RevModPhys.88.035005,
PhysRevLett.121.086803, PhysRevLett.123.246801, PhysRevLett.121.026808}. The
$Q$ matrix is defined by
\begin{equation*}
Q(\beta )=|\tilde{u}_{\text{R}}(\beta )\rangle \langle \tilde{u}_{\text{L}%
}(\beta )|-|u_{\text{R}}(\beta )\rangle \langle u_{\text{L}}(\beta )|,
\end{equation*}%
where the right vector $|u_{\text{R}}\rangle $ and left vector $|u_{\text{L}%
}\rangle $ are defined by
\begin{equation*}
\check{h}_{\tilde{k}}|u_{\text{R}}\rangle =E(\tilde{k})|u_{\text{R}}\rangle
,\quad \check{h}_{\tilde{k}}^{\dag }|u_{\text{L}}\rangle =E^{\ast }(\tilde{k}%
)|u_{\text{L}}\rangle .
\end{equation*}%
$|\tilde{u}_{\text{R}}\rangle \equiv \sigma _{z}|u_{\text{R}}\rangle $ and $|%
\tilde{u}_{\text{L}}\rangle \equiv \sigma _{z}|u_{\text{L}}\rangle $ are
also right and left eigenvectors, with eigenvalues $-E$ and $-E^{\ast }$ due
to the chiral symmetry. The normalization conditions are $\langle u_{\text{L}%
}|u_{\text{R}}\rangle =\langle \tilde{u}_{\text{L}}|\tilde{u}_{\text{R}%
}\rangle =1$ and $\langle u_{\text{L}}|\tilde{u}_{\text{R}}\rangle =\langle
\tilde{u}_{\text{L}}|u_{\text{R}}\rangle =0$. The \textquotedblleft $Q$
matrix is off-diagonal, namely $Q=q\sigma _{+}+q^{-1}\sigma _{-}$ where $q=%
\sqrt{h_{2}\left( \tilde{k}\right) /h_{1}\left( \tilde{k}\right) }\exp
\left\{ i\left[ \theta _{2}\left( \tilde{k}\right) -\theta _{1}\left( \tilde{%
k}\right) \right] /2\right\} $. The non-Bloch winding number is given by
\begin{equation}
\nu _{Q}=\frac{i}{2\pi }\int_{C_{\beta }}q^{-1}dq
\label{the winding number of Q matrix}
\end{equation}%
which is difference of winding numbers of two winding vectors $h_{+}\left(
\tilde{k}\right) e^{i\theta _{+}\left( \tilde{k}\right) }$ and $h_{-}\left(
\tilde{k}\right) e^{i\theta _{-}\left( \tilde{k}\right) }$.

According to the geometry of two winding vectors $h_{+}\left( \tilde{k}%
\right) e^{i\theta _{+}\left( \tilde{k}\right) }$ and $h_{-}\left( \tilde{k}%
\right) e^{i\theta _{-}\left( \tilde{k}\right) }$, at the phase transitions
points, the quantums must meet the relationship:%
\begin{eqnarray}
\bar{t}_{1} &=&t_{2}e^{i\phi }+t_{3}e^{-3i\phi },\text{ }
\label{phase transitions points} \\
\text{ }\bar{t}_{1} &=&t_{2}e^{-i\phi }+t_{3}e^{3i\phi }.  \notag
\end{eqnarray}%
Taking $t_{2}=1$ and $\gamma =4/3$ for example, the phase transition points
are $t_{1}=1.5660$ and $t_{1}=1.7050$ according to the Eq. (\ref{phase
transitions points}). At $t_{1}=1.5660 $, the system transforms firstly from
topological nontrivial phase to topological trivial phase which is the
result in Ref. \cite{PhysRevLett.121.086803}.

To summarize the approach: With a similarity transformation, the
non-Hermitcity of the model is transformed from $t_{1}$ term to $t_{2}$ and $%
t_{3}$ terms. According the non-Hermitcity of the equivalent model, the
imaginary gauge field $\Phi $ is the obtained. Using the imaginary gauge
field, the partner model is constructed with the Peierls replacement $\tilde{%
k}\rightarrow k+i\Phi $. At last, the non-Hermitian winding number is solved.

\section{Two applications}

\subsection{The non-Hermitian SSH model with spin-orbit coupling}
\label{The non-Hermitian SSH model with spin-orbit coupling}

When spin-orbit coupling is taking into account, the non-Hermitian version
of SSH model can be written by the Hamiltonian $H=H_{\mathrm{SSH}}+H_{%
\mathrm{SOC}}$ \cite{HAN201968} where
\begin{eqnarray*}
H_{\mathrm{SSH}}\! &=&\sum\limits_{n,\sigma }\left[ (t-\delta )a_{n,\sigma
}^{\dagger }b_{n,\sigma }+(t+\delta )b_{n,\sigma }^{\dagger }a_{n,\sigma
}\right. \\
&&\left. ~~~+t^{\prime }a_{n+1,\sigma }^{\dagger }b_{n,\sigma }+t^{\prime
}a_{n,\sigma }^{\dagger }b_{n+1,\sigma }\right] ,
\end{eqnarray*}%
where $a_{n,\sigma }^{\dagger }(a_{n,\sigma })$ and $b_{n,\sigma }^{\dagger
} $ ($b_{n,\sigma }$) are the electron creation (annihilation) operators
with spin $\sigma =(\uparrow \text{or}\downarrow )$ on the sublattices $A$
and $B$ of the \textit{n}th unit cell, respectively. The non-Hermiticity of
the Hamiltonian is due to the introduction of $\delta $.
The spin-orbit coupling Hamiltonian is described by
\begin{equation*}
H_{\mathrm{SOC}}=\sum\limits_{n,\sigma }[\lambda a_{n,\sigma }^{\dagger
}b_{n,-\sigma }-\lambda ^{\prime }a_{n,\sigma }^{\dagger }b_{n-1,-\sigma }+%
\mathrm{h.c.}],
\end{equation*}%
where $\lambda $ and $\lambda ^{\prime }$ denote the spin-orbit coupling
amplitudes in the unit cell and between two adjacent unit cells,
respectively. When the spin-orbit coupling is Dresshaus type, the coupling
amplitudes $\lambda $ and $\lambda ^{\prime }$ are real value. Otherwise the
coupling amplitudes are imaginary value when the spin-orbit coupling is
Rashba type. Adopting periodic boundary conditions and Fourier transforming,
the Hamiltonian $H=H_{\mathrm{SSH}}+H_{\mathrm{SOC}}$ can be easily written
as $H=\sum_{k}\psi _{k}^{\dagger }h_{k}\psi _{k}$, where $\psi _{k}^{\dagger
}=(a_{k,\uparrow },b_{k,\uparrow },a_{k,\downarrow },b_{k,\downarrow
})^{\dagger }$, and
\begin{equation}
h_{k}=\left(
\begin{array}{cccc}
0 & \zeta _{k}-\delta & 0 & \varsigma _{k} \\
\zeta _{k}^{\ast }+\delta & 0 & \varsigma _{k}^{\ast } & 0 \\
0 & \varsigma _{k} & 0 & \zeta _{k}-\delta \\
\varsigma _{k}^{\ast } & 0 & \zeta _{k}^{\ast }+\delta & 0%
\end{array}%
\right) ,  \label{hk-matrix}
\end{equation}%
with $\zeta _{k}=t+t^{\prime }e^{-ik}\ $and$\quad \varsigma _{k}=\lambda
-\lambda ^{\prime }e^{-ik}$.

The non-Hermitian skin effect of the model can be analysed as follow. When
the spin-orbit coupling effect is negligible, the Hamiltonian in Eq. (\ref%
{hk-matrix}) is the direct-sum of two non-Hermitian SSH Hamiltonians. The
non-Hermitian skin effect exists in the system. When the spin-orbit coupling
effect dominates the system, the Hamiltonian (\ref{hk-matrix}) is a
Hermitian. Therefore, the non-Hermitian skin effect exists in system.
Considering the spin-up and spin-down of the particles, two channels have
been provided for the particles tunneling from A sites to B sites of the AB
sublattice. The spin-orbit coupling effect provides a new channel which
induces the different spin particles tunneling from A sites to B sites.

Referring the non-Hermitian terms $t_{1}\pm \delta =\bar{t}_{1}e^{\pm \phi }$%
, the asymmetric hopping is equivalent to an imaginary magnetic field
applying with the imaginary vector potential $-i\phi $. Considering the two
channels, the effective imaginary field with vector potential $-2i\phi $ is
applying to the lattice. The $-2i\phi $ vector potential suggests the wave
vector should be with the Peierls replacement $\tilde{k}\rightarrow k+2i\phi
$.

To finish this replacement, we do the similarity transformation to the
Hamiltonian in Eq. (\ref{hk-matrix}) $\tilde{h}_{k}=S_{4}^{-1}h_{k}S_{4}$
with a diagonal matrix $S$ whose diagonal elements are $\left\{ e^{-2\phi
},1,e^{-2\phi },1\right\} $. After the similarity transformation, we obtain
the partner model
\begin{equation}
\tilde{h}_{k}=\left(
\begin{array}{cccc}
0 & \tilde{\zeta}_{k,+} & 0 & \varsigma _{k,+} \\
\tilde{\zeta}_{k,-} & 0 & \varsigma _{k,-} & 0 \\
0 & \varsigma _{k,+} & 0 & \tilde{\zeta}_{k,+} \\
\varsigma _{k,-} & 0 & \tilde{\zeta}_{k,-} & 0%
\end{array}%
\right)  \label{the partner of SOC SSH}
\end{equation}%
here $\tilde{\zeta}_{k,\pm }=te^{\pm \phi }+t^{\prime }e^{\pm 2\phi }e^{\mp
ik}\ $and$\quad \tilde{\varsigma}_{k,\pm }=e^{\pm 2\phi }\left( \lambda
-\lambda ^{\prime }e^{\mp ik}\right) $. The wave function becomes $\psi
_{k}^{\dagger }=\left( e^{2\phi }a_{k,\uparrow },b_{k,\uparrow },e^{2\phi
}a_{k,\downarrow },b_{k,\downarrow }\right) ^{\dag }$. The partner is still
an non-Hermitian model. Under the OBC, the partner can be obtained by a
similarity transformation $\tilde{H}=S_{1}^{-1}HS_{1}$ with a diagonal
matrix $S_{1}$ given in Eq. (\ref{the transformation matrix S1}), where $%
r_{1}=\left\vert \left( t_{1}-\delta \right) /\left( t_{1}+\delta \right)
\right\vert $. Under the similarity transformation, the non-Hermitcity is
transformed to $t$ and $\lambda $ terms and the $t_{2}$ term remains
unchanged. The wavefunction $|\psi \rangle =(a_{1},b_{1},a_{2},b_{2},\cdots
,a_{L},b_{L})^{T}\ $\ becomes $|\bar{\psi}\rangle =S_{1}^{-1}|\psi \rangle $.

For the non-Hermitian terms $\lambda e^{\pm 2\phi }$, the asymmetric
hoppings are equivalent to an imaginary magnetic field applying to the
sublattice with the imaginary vector potential $2i\phi $. Considering the
spin-up and spin-down channels, the effective imaginary field with vector
potential is $-2i\phi $ applying the lattice. The effective imaginary fields
are canceled. It indicates\ that the non-Herimitian skin effect disappears
and the wave vector is a good quantum number in the partner model in Eq. (%
\ref{the partner of SOC SSH}).

The Hamiltonian $\tilde{h}_{k}$ in Eq. (\ref{hk-matrix}) can be brought into
the direct-sum of two block diagonal form by the unitary transformation
\begin{equation}
\hat{h}_{k}=U^{\dag }\tilde{h}_{k}U=\hat{h}_{\text{up}}\oplus \hat{h}_{\text{%
down}}  \label{the unitary transformation of the partner of SOC SSH}
\end{equation}%
here the two block matrixes are%
\begin{eqnarray}
\hat{h}_{\text{up}} &=&\left( \tilde{\zeta}_{k,-}+\tilde{\varsigma}%
_{k,-}\right) \sigma _{+}+\left( \tilde{\zeta}_{k,+}+\tilde{\varsigma}%
_{k,+}\right) \sigma _{-}  \label{Block up of the partner of SOC SSH} \\
&=&e^{-2\phi }\left( \eta _{1}^{\ast }+\delta \right) \sigma _{+}+e^{2\phi
}\left( \eta _{1}-\delta \right) \sigma _{-},  \notag
\end{eqnarray}%
\begin{eqnarray}
\hat{h}_{\text{down}}\left( \tilde{k}\right)  &=&\left( \tilde{\zeta}_{k,-}-%
\tilde{\varsigma}_{k,-}\right) \sigma _{+}+\left( \tilde{\zeta}_{k,+}-\tilde{%
\varsigma}_{k,+}\right) \sigma _{-}
\label{Block down of the partner of SOC SSH} \\
&=&e^{-2\phi }\left( \eta _{2}^{\ast }+\delta \right) \sigma _{+}+e^{2\phi
}\left( \eta _{2}-\delta \right) \sigma _{-}.  \notag
\end{eqnarray}%
here $\eta _{1}=\zeta _{k}+\varsigma _{k}$ and $\eta _{2}=\zeta
_{k}-\varsigma _{k}$. The unity matrix is
\begin{equation*}
U=\frac{1}{\sqrt{2}}\left(
\begin{array}{cccc}
0 & 1 & 0 & 1 \\
1 & 0 & 1 & 0 \\
0 & 1 & 0 & -1 \\
1 & 0 & -1 & 0%
\end{array}%
\right) .
\end{equation*}
$\hat{h}_{\text{up}}$\ and $\hat{h}_{\text{down}}$ show chiral symmetry
defined respectively as $\mathcal{C}\hat{h}_{\text{up(down)}}\mathcal{C}%
^{-1}=-\hat{h}_{\text{up(down)}}$, where $\mathcal{C=\sigma }_{z}$. The
topological invariants of the non-Hermitian systems in Eq. (\ref{Block up of
the partner of SOC SSH}) and (\ref{Block down of the partner of SOC SSH})
are the winding numbers of energy defined in Eq. (\ref{the winding number of
energy}). The winding number $\hat{h}_{\text{up}}$ in Eq. (\ref{Block up of
the partner of SOC SSH}) is just the summation of the winding numbers of the
two vectors $\left( \eta _{1}-\delta \right) $, $-\left( \eta _{1}^{\ast
}+\delta \right) $. The winding number $\hat{h}_{\text{down}}$ in Eq. (\ref%
{Block down of the partner of SOC SSH}) is the summation of the winding
numbers of the two vectors $\left( \eta_{2}-\delta \right) $,
 $-\left( \eta _{2}^{\ast }+\delta \right) $. The
results have obtained and comfirmed mumerically in Ref. \cite{HAN201968}.

\subsection{The topological defect on non-Hermitian SSH model}

\label{The topological defect on non-Hermitian SSH model}

The above method can also been applied to other model. For example, the
topological defect on non-Hermitian SSH model. The defect is an useful tool
to probe the nontrivial topological properties of bulk systems \cite%
{PhysRevB.82.115120, PhysRevB.93.035134, Lu_2011, Lang_2014}. The
topological defect on non-Hermitian lattices with spatially distributed gain
and loss was discussed \cite{PhysRevA.97.042118, PhysRevA.99.062107,
PhysRevB.98.094307}. Two bulks with topologically distinct phases contacting
through a common boundary form a domain wall configuration is a special
topological defect where the domain wall state is exponentially localized at
the interface between two different bulks. Recently, some non-Hermitian
topological systems in domain wall configuration were studied theoretically
and experimentally \cite{PhysRevLett.115.200402, 2020NatPh..16..761X,
2020NatPh..16..747H, PhysRevB.100.035102, PhysRevB.98.094307}.

A non-Hermitian SSH model in a domain-wall configuration on a ring has been
investigated numerically in Ref. [\onlinecite{PhysRevB.100.035102}]. When
the length of the two chains is large, the domain can be taken as a
perturbation. In particular, the coupling of the domain states can also be
neglected. The energy spectra of the domain-wall configuration on the ring
can be well approximated by that on a chain. A non-Hermitian SSH model in a
domain-wall configuration on a chain is pictorially shown in Fig. \ref%
{Non_Hermit_SSH_domain}. The Hamiltonian can be written as
\begin{equation}
H=H_{L}+H_{R},
\label{A non-Hermitian SSH model in a domain-wall configuration}
\end{equation}%
where
\begin{align*}
H_{\alpha }& =\sum_{j}\left[ t_{\alpha }e^{\phi _{\alpha }}a_{j}^{\dagger
}b_{j}+t_{\alpha }e^{-\phi _{\alpha }}b_{j}^{\dagger }a_{j}\right. \\
& \left. +t_{2}a_{j+1}^{\dagger }b_{j}+t_{2}b_{j}^{\dagger }a_{j+1}\right],
\end{align*}%
here $\alpha =(L,R)$ denotes the left or right bulk. $a_{j}^{\dagger }$ ($%
a_{j}$) and $b_{j}^{\dagger }$ ($b_{j}$) are the creation (annihilation)
operators for the sublattice sites $a$ and $b$ on the $j$-th unit cell. The
left and right bulks have different parameters and contain $N_{L}$ and $%
N_{R} $ unit cells respectively. The non-Hermicity of the system is from the
difference intra-cell hopping with phase factor of amplification/attenuation
$e^{\pm \phi _{\alpha }}$. The non-Hermitian SSH model has also chiral
symmetry $\Gamma H\Gamma ^{-1}=-H$, with the chiral-symmetry operator $%
\Gamma =\sum_{j=1}^{N_{L}+N_{R}}(a_{j}^{\dagger }a_{j}-b_{j}^{\dagger
}b_{j}) $.

\begin{figure}[tbp]
\begin{center}
\includegraphics[width=8cm]{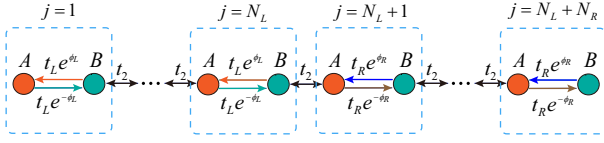}
\end{center}
\caption{Schematic of a non-Hermitian SSH model under the domain-wall
configuration. The intra-cell hopping from sublattice sites $a$ to $b$ in
the left(right) bulk is $t^{L(R)}e^{\protect\phi_{L(R)}}$, while the
intra-cell hopping from $b$ to $a$ is $t^{L(R)}e^{-{\protect\phi_{L(R)}}}$.
For simplicity, we assume both bulks have the same $t_2$. }
\label{Non_Hermit_SSH_domain}
\end{figure}

To understand the topological properties of the model, we do the following
transformation to the Hamiltonian in Eq. (\ref{A non-Hermitian SSH model in
a domain-wall configuration}):%
\begin{equation*}
H=\left\langle \Psi \right\vert h\left\vert \Psi \right\rangle =\left\langle
\Phi \right\vert \bar{h}\left\vert \Phi \right\rangle ,
\end{equation*}%
here $\bar{h}=S^{-1}hS$ and%
\begin{equation*}
|\psi \rangle =(a_{1},b_{1},\cdots ,a_{N_{L}},b_{N_{L}},,\cdots
,a_{N_{L}+N_{R}},b_{N_{L}+N_{R}})^{T}
\end{equation*}%
and
\begin{eqnarray*}
\left\vert \Phi \right\rangle &=&S|\psi \rangle \\
&=&(c_{1},d_{1},\cdots ,c_{N_{L}},d_{N_{L}},,\cdots
,c_{N_{L}+N_{R}},d_{N_{L}+N_{R}})^{T}.
\end{eqnarray*}%
The transformation matrix $S=S_{1}S_{2}$. $S_{1}$ and $S_{2}$ are the
diagonal matrices. The diagonal elements of $S_{1}$ are seted as
\begin{eqnarray*}
&&\left\{ 1,r,r,r^{2},r^{2},\cdots ,r^{N_{L}-1},r^{N_{L}-1},r^{N_{L}}\right.
, \\
&&\left. r^{N_{R}},r^{N_{R}-1},r^{N_{R}-1},\cdots ,r^{2},r^{2},r,r,1\right\}
\end{eqnarray*}%
with $r=e^{\left( \phi _{R}-\phi _{L}\right) /2}$. After the similarity
transformation $S_{1}$, we have the same amplification/attenuation $e^{\pm
\phi }$ of phase factor in asymmetric couplings of the two bulks, here $\phi
=\left( \phi _{R}+\phi _{L}\right) /2$. Then further doing the similarity
transformation $S_{2}$ with a diagonal matrix whose diagonal elements are
selected as
\begin{equation*}
\left( 1,r,r,r^{2},r^{2},\cdots ,r^{N_{L}},r^{N_{L}},\cdots
,r^{N_{L}+N_{R}-1},r^{N_{L}+N_{R}-1},r^{N_{L}+N_{R}}\right) ,
\end{equation*}%
we get the Hermitian Hamiltonian of the two bulks connecting in a chain
configuration
\begin{equation*}
H_{\alpha }=\sum_{j}\left( t_{\alpha }c_{j}^{\dagger }d_{j}+t_{\alpha
}d_{j}^{\dagger }c_{j}+t_{2}c_{j+1}^{\dagger }d_{j}+h.c.\right).
\end{equation*}%
When the two bulks connecting in a ring, we get the partner model of
non-Hermitian SSH model in domain configuration studied in Ref.\cite%
{PhysRevB.100.035102}. The Bloch winding number of this model is the winding
number different $\nu _{R}-\nu _{L}$ of the two bulks. The transition points
of the two bulks are $t_L=t_{2}$ and $t_R=t_{2}$.

Another type of defect is impurity which can vary the forward and backward
scattering amplitude of the continuous states and induce gap bound states.
The gap bound states are exponentially localized at the impurity \cite%
{Schomerus:13, Longhi:14}. A hard-wall boundary is a special impurity which
is equivalent to the infinity impurity strength and the forward scattering
is forbidden. The single-impurity problem in the non-Hermitian SSH model is
shown in Ref. [\onlinecite{PhysRevB.102.075404}].

According to bulk-boundary correspondence, when the lattice is long enough,
the impurity state on the dimer ring can be approximated by that on the long
chain since the boundary scattering can be regarded as a perturbation. The
Hamiltonian of impurity problem is described by%
\begin{equation}
H_{ip}=H_{NSSH}+va_{0}^{\dag }a_{0}=\psi ^{\dagger }h_{ip}\psi ,
\label{impurity on the non-Hermitian SSH lattice}
\end{equation}%
where $v$ is the strength of the impurity potential and $\psi ^{\dagger
}=\left( a_{-L/2}^{\dag },b_{-L/2}^{\dag },a_{-L/2+1}^{\dag
},b_{-L/2+1}^{\dag }\cdot \cdot \cdot ,a_{L/2}^{\dag },b_{L/2}^{\dag
}\right) $ with lattice length $L$. The non-Hermitian SSH Hamiltonian
\begin{equation*}
H_{NSSH}=\sum_{j=-L/2}^{L/2}\left( te^{\phi }a_{j}^{\dagger }b_{j}+te^{-\phi
}b_{j}^{\dagger }a_{j}+t^{\prime }a_{j+1}^{\dagger }b_{j}+t^{\prime
}b_{j+1}^{\dagger }a_{j}\right) .
\end{equation*}%
Here $te^{\pm \phi }$ is the right (left) intra-hopping amplitude and $%
t^{\prime }$\ is the inter-hopping amplitude.

The partner of the impurity model can be obtained by a conventional
similarity transformation $\tilde{h}_{ip}=S^{-1}h_{ip}S$ with a diagonal
matrix
\begin{equation*}
S=\text{diag}\{1,r,r,r^{2},r^{2},\cdots ,r^{L/2},r^{L/2}\},
\end{equation*}%
here $r=e^{\phi }$. After this transformation. $H_{NSSH}$ becomes the
standard SSH model with the intra-hopping amplitude and the inter-hopping
amplitude $t^{\prime }$. The wavefunction $|\tilde{\psi}\rangle =S^{-1}|\psi
\rangle $ and the the strength of the impurity potential $v$ remains
unchanged. The transition points are $t=t^{\prime }$.

\section{Summary}

\label{Summary}

In summary, we have proposed a way to construct a counterpart of
non-Hermitian SSH model. The kernel of this method is to find the
effectively imaginary gauge field referring to the non-Hermitian skin effect
under PBC. Under the OBC, the imaginary magnetic flux vanishes and the OBC
spectra is used to approximate the PBC spectra. The corresponding
Hamiltonian is the counterpart of the original model. In fact, the partner
model can be obtained by a similarity transformation. Due to the skin effect
is eliminated from the wave vector, we can study the topological equivalent
model in conventional Bloch space. Our work gives an alternating method to
study the non-Hermitian SSH model with its counterpart. Several
non-Hermitian SSH models are used to illustrate the method and phase
transition points are given in analytic form. This method is expected to be
used to construct the counterpart of a class of non-Hermitian model. In view
of the non-reciprocal hopping, the similarity transformations and the
effective imaginary gauge fields may depend on model details without any
general rules. Not all of the non-Hermitian models may have the partner
models constructed by this method, for example, the models proposed in Refs.
\cite{PhysRevLett.125.126402, PhysRevLett.125.186802, PhysRevB.101.020201}.
Finding their counterparts is still a challenging subject.

\begin{acknowledgments}
This work was supported by Hebei Provincial Natural Science Foundation of
China (Grant No. A2012203174, No. A2015203387) and National Natural Science
Foundation of China (Grant No. 10974169, No. 11304270).
\end{acknowledgments}

%\bibliography{Ref}


\end{document}